\documentclass[11pt,a4paper]{article}
\pdfoutput=1

\usepackage{jheppub}
\usepackage{latexsym}
\usepackage{slashed}
\usepackage{multirow}
\usepackage{color}
\usepackage[usenames,dvipsnames,table]{xcolor}
\usepackage{graphicx}% Include figure files
\usepackage{epsfig}  % Include figure files
\usepackage{dcolumn}% Align table columns on decimal point
\usepackage{bm}% bold math
\usepackage{dcolumn}% Align table columns on decimal point
\usepackage{textcomp}% Align table columns on decimal point
\usepackage{float}
\usepackage{subfigure}
\usepackage[]{hyperref}
\usepackage{makecell}
\usepackage{color}
\usepackage{pifont}
 \usepackage{caption}
  \newcommand{\be}{\begin{equation}}
   \newcommand{\ee}{\end{equation}}

\hypersetup{
  bookmarks=true,         % show bookmarks bar?
  unicode=false,          % non-Latin characters in Acrobat?s bookmarks
  pdftoolbar=true,        % show Acrobat?s toolbar?
 pdfmenubar=true,        % show Acrobat?s menu?
 pdffitwindow=true,     % window fit to page when opened
 pdfstartview={FitH},    % fits the width of the page to the window
 pdfsubject={Dark matter searches Neutrino Phenomenology},   % subject of the document
 pdfnewwindow=true,      % links in new window
 pdfcreator={RevTeX},
 colorlinks=true,       % false: boxed links; true: colored links
 linkcolor=red,          % color of internal links
 citecolor=blue,        % color of links to bibliography
 filecolor=black,      % color of file links
 urlcolor=blue,           % color of external links
  }

\title{Constraints on dark matter annihilation to fermions and a photon}

\author[a,b,c]{Debtosh Chowdhury,}
\emailAdd{Debtosh.Chowdhury@polytechnique.edu}
\author[d]{Abhishek M. Iyer,}
\emailAdd{iyera@na.infn.it}
\author[e]{ and Ranjan Laha.}

\affiliation[a]{Istituto Nazionale di Fisica Nucleare,
Sezione di Roma, Piazzale Aldo Moro 2, I-00185 Roma, Italy}

\affiliation[b]{Centre de Physique Th{\'e}orique, {\'E}cole Polytechnique, F-91128 Palaiseau Cedex, France}

\affiliation[c]{Laboratoire de Physique Th{\'e}orique, Universit{\'e} Paris-Sud, F-91405 Orsay Cedex, France}
\affiliation[d]{INFN-Sezione di Napoli, 
Via Cintia, 80126 Napoli, Italia}
\affiliation[e]{Kavli Institute for Particle Astrophysics and Cosmology (KIPAC), Department of Physics, Stanford University, Stanford, CA 94305, USA and SLAC National Accelerator Laboratory, Menlo Park, CA 94025, USA} 
\affiliation[e]{PRISMA Cluster of Excellence and
             Mainz Institute for Theoretical Physics,
             Johannes Gutenberg-Universit\"{a}t Mainz, 55099 Mainz, Germany}

\emailAdd{ranjalah@uni-mainz.de}

\abstract{We consider Majorana dark matter annihilation to fermion - anti-fermion pair and a photon in the effective field theory paradigm, by introducing dimension 6 and dimension 8 operators in the Lagrangian. For a given value of the cut-off scale, the latter dominates the annihilation process for heavier dark matter masses.   We find a cancellation in the dark matter annihilation to a fermion - anti-fermion pair when considering the interference of the dimension 6 and the dimension 8 operators.  Constraints on the effective scale cut-off is derived while considering indirect detection experiments and the relic density requirements and they are compared to the bound coming from collider experiments.}

\begin{document}
%\preprint{IP/BBSR/2017-7}
\maketitle
%\flushbottom
\newpage
%%%%%%%%%%%%%%%%%%%%%%%%%%%%%%%%%%%%%%%%%%%%%%%%%%%%%%%
%%%%%%%%%%%%%%%%%%%%%%%%%%%%%%%%%%%%%%%%%%%%%%%%%%%%%%%

\section{Introduction}
\label{sec:Introduction}

Astrophysical observations at all scales confirm the presence of dark matter~\cite{Strigari:2013iaa}.  The ubiquitous astrophysical discovery of dark matter from the Galactic scale to the cosmic microwave background scale confronts us with many profound mysteries of nature~\cite{Jungman:1995df,Bergstrom:2000pn,Feng:2010gw,Bergstrom:2012fi,Klasen:2015uma}.  Although astrophysical observations give us important clues on dark matter properties, precise questions about it can only be answered once we determine the particle properties of dark matter.
	
Numerous dark matter particle candidates exist with masses ranging from $\sim$ 10$^{-22}$ eV to 10$^{2}$ M$_\odot$, but perhaps the most widely searched for particle goes under the generic name of weakly interacting massive particles (WIMPs).  WIMPs have masses in between a few GeV to few tens of TeV, and interact with the Standard Model particles with ``weak" strength~\cite{Gunn:1978gr,Stecker:1978du,Zeldovich:1980st}.  Dark matter searches are conducted in colliders~\cite{Aad:2014vea}, indirect detection~\cite{Bringmann:2012ez,Danninger:2014xza,Buckley:2013bha,Ackermann:2015zua,Aartsen:2013dxa,Conrad:2015bsa,Porter:2011nv,Dasgupta:2012bd,Laha:2012fg,Ng:2013xha,Murase:2015gea,Powell:2016zbo,Khatun:2017adx, Vogel:2017fmc}, and direct detection~\cite{Peter:2013aha,Lewin:1995rx,Freese:2012xd,Cushman:2013zza,Akerib:2013tjd,Anand:2013yka,Catena:2014epa,Panci:2014gga,Laha:2013gva,Laha:2015yoa,Laha:2016iom}.

In indirect detection of dark matter, we search for faint signals of dark matter annihilation and decay from the cosmos amidst the overwhelming conventional astrophysical background.  These searches can give us information about their properties which are not easily accessible otherwise~\cite{Ng:2013xha}.  Due to the enormous and varied astrophysical background, it is often necessary to either search for a signal unique to the dark matter particle candidate or device clever search strategies~\cite{Speckhard:2015eva}.

Dark matter interactions to the Standard Model sector can be parametrized by higher dimensional non-renormalizable operators using the effective field theory technique \cite{Beltran:2008xg,Beltran:2010ww,Goodman:2010ku,Goodman:2010yf,Goodman:2010qn,Harnik:2008uu,Cao:2009uw,Cheung:2012gi,DeSimone:2013gj,Bell:2015sza,Bell:2013wua,Carpenter:2015xaa}. These operators are suppressed by different powers of the effective cut-off scale $\Lambda$. If the new physics is sufficiently decoupled from the Standard Model particles then this is an adequate description.  There have been recent discussion about the validity of effective field theory, especially at the colliders~\cite{Shoemaker:2011vi,Busoni:2013lha,Busoni:2014sya,Busoni:2014haa}.  In spite of its limitations, effective field theory approach to dark matter is useful to classify various interactions and make progress in  understanding dark matter physics.

Various well motivated new physics extension of the Standard Model predict that the dark matter particle is a Majorana fermion.  General considerations suggest that the annihilation of Majorana dark matter particles to fermions in the $s$-wave channel is proportional to $(m_f/m_\chi)^2$, where $m_f$ and $m_\chi$ stands for the mass of the Standard Model fermion and dark matter particle respectively~\cite{Goldberg:1983nd,Weiler:2013ama,Weiler:2013hh}.  The rate in the $p$-wave channel is proportional to $v^2$, where $v$ is the relative velocity of two incoming dark matter particles.  

Given that the dark matter particle is expected to have a mass $\gtrsim$ 100 GeV, and almost all the Standard Model fermions (except the top quark) have a mass of $<$ 5 GeV, this implies a huge suppression of the $s$-wave annihilation rate.  The dark matter velocity in the Solar circle is $\sim $10$^{-3}$c, and the typical velocities in clusters and dwarf galaxies are $\sim$10$^{-2}$c and $\sim$10$^{-4}$c respectively.  These small velocities ensure that the $p$-wave contribution to non-relativistic Majorana dark matter particles annihilating to fermions is velocity suppressed.

A possible way out of this conundrum was suggested long ago in which a photon is radiated out from the final state fermion or the charged mediator and this lifts the suppression~\cite{Bergstrom:1989jr,Flores:1989ru}.  This contribution has also been included in numerical packages which calculate dark matter properties for supersymmetric dark matter candidates~\cite{Bringmann:2007nk}.  Recently it has also been realized that in general the radiation of electroweak gauge bosons lift this suppression~\cite{Bell:2008ey,Bell:2010ei,Bell:2011eu,Bell:2011if,Ciafaloni:2011gv,Ciafaloni:2011sa,Garny:2011cj,Bringmann:2013oja,Barger:2011jg,Garny:2013ama,Kopp:2014tsa,Okada:2014zja,Garny:2015wea}.  Additional work has also been done regarding gluon~\cite{Garny:2011ii,Bringmann:2015cpa} and Higgs radiation~\cite{Luo:2013bua}.

It is tempting to ask if it is possible to incorporate boson bremsstrahlung in the framework of effective field theory of dark matter.  Ref.~\cite{DeSimone:2013gj} presented a list of operators which describes Majorana dark matter annihilation to fermion - anti-fermion pair and an electroweak gauge boson. It was shown that dimension-8 operator is required to consistently describe a Majorana dark matter annihilation into a  fermion - anti-fermion pair and an electroweak gauge boson.

In this present work, using these operators, constraints on $\Lambda$ are obtained from the relic density requirement and present limits on dark matter annihilation is compared with those obtained from collider.    The operators that we consider do not have a non-relativistic limit and hence the constraints from dark matter direct detection experiments do not apply~\cite{Cheung:2012gi}. 

In Section~\ref{sec:effective operator model} we discuss the effective operators and compute the cutoff scale $\Lambda$ to obtain the correct dark matter relic density.  We compare these values with the constraints from colliders and indirect detection experiments.  We conclude in Section~\ref{sec:conclusions}.

%%%%%%%%%%%%%%%%%%%%%%%%%%%%%%%%%%%%%%%%%%%%%%%%%%%%%%%%%%%%%%%%%%%%%%%%%%%%%%%%%%%
\section{Effective operator model}
\label{sec:effective operator model}
%%%%%%%%%%%%%%%%%%%%%%%%%%%%%%%%%%%%%%%%%%%%%%%%%%%%%%%%%%%%%%%%%%%%%%%%%%%%%%%%%%

We assume that the dark matter particle is a Majorana fermion and is represented by $\chi$.  The lowest order interaction between dark matter particles and Standard Model fermions, denoted by $f$, that we consider is of the form~\cite{DeSimone:2013gj}
\begin{eqnarray}
\mathcal{L}_{\rm d = 6} = \dfrac{1}{\Lambda^2} (\bar{\chi} \gamma^5 \gamma^\mu \chi)(\bar{f} \gamma_\mu f) \, .
\label{eq:dim6}
\end{eqnarray}
The higher order Lagrangian that we consider is of the form~\cite{DeSimone:2013gj}
\begin{eqnarray}
\mathcal{L}_{\rm d = 8} &=& \dfrac{1}{\Lambda^4} (\bar{\chi} \gamma^5 \gamma^\mu \chi)\bigg[\bigg(\bar{f}_L \overleftarrow{D_\rho} \bigg)\gamma_\mu \bigg( \overrightarrow{D^\rho} f_L \bigg) + \bigg(\bar{f}_R \overleftarrow{D_\rho} \bigg)\gamma_\mu \bigg( \overrightarrow{D^\rho} f_R \bigg) \bigg] \, ,
\label{eq:dim8}
\end{eqnarray}
where following Ref.~\cite{DeSimone:2013gj}, we define
\begin{eqnarray}
&&\bar{f}_L \overleftarrow{D}_\mu = (\partial_\mu \bar{f}_L) - i g \dfrac{\sigma^i}{2} W_\mu ^i \bar{f}_L - i g' Y_f B_\mu \bar{f}_L \,, \\
&&\overrightarrow{D}_\mu f_L = (\partial_\mu f_L) + i g \dfrac{\sigma^i}{2} W_\mu ^i f_L + i g' Y_f B_\mu f_L  \,, \\
&&\bar{f}_R \overleftarrow{D}_\mu = (\partial_\mu \bar{f}_R) - i g' Y_f B_\mu \bar{f}_R \,, \\
&& \overrightarrow{D}_\mu f_R = (\partial_\mu f_R) + i g' Y_f B_\mu f_R \,, \\
\label{eq:overleftarrow overrightarrow}
\end{eqnarray}
where $\sigma^i$ denotes the Pauli matrices, $W^\mu$ and $B^\mu$ denote the Standard Model gauge bosons, $Y_f$ denotes the hypercharge and $g$ and $g'$ denote the Standard Model SU(2) and U(1) gauge couplings respectively.

The total Lagrangian is given by 
\begin{eqnarray}
\mathcal{L} = d_6 \, \mathcal{L}_{\rm d = 6} + d_8 \, \mathcal{L}_{\rm d = 8} \, ,
\label{eq:total lagrangian}
\end{eqnarray}
where $d_6$ and $d_8$ are arbitrary complex constants.

\subsection{$\chi + \chi \rightarrow f + \bar{f}$}
We will first calculate the cross section for the process $\chi (k_1) + \chi (k_2) \rightarrow f(p_1) + \bar{f}(p_2)$, where $k_i$ and $p_i$ denote the four momenta.  An important ingredient in our calculation is that we take into account both the dimension 6 and dimension 8 terms.  As will be seen due to the interference of these two terms, there is a cancellation feature in the $\sigma v$ which is not present when one only considers dimension 6 term.  From Eq.~\eqref{eq:total lagrangian}, the relevant part of the Lagrangian responsible for the process is
\begin{align}
\mathcal{L} &= \dfrac{d_6}{\Lambda^2} (\bar{\chi} \gamma^5 \gamma^\mu \chi)(\bar{f} \gamma_\mu f) 
+ \dfrac{d_8}{\Lambda^4} (\bar{\chi} \gamma^5 \gamma^\nu \chi) \bigg[\left(\partial_\rho \bar{f} \right)\gamma_\nu \left(\partial^\rho f \right) \bigg] \, .
\label{eq:relevant lagrangian for 2 body final state}
\end{align}
From this Lagrangian, we obtain the following cross section for the process $\chi (k_1) + \chi (k_2) \rightarrow f(p_1) + \bar{f}(p_2)$
\begin{align}
\sigma = \dfrac{1}{v} \dfrac{v^2}{3\pi \Lambda^8} &\sqrt{1-\dfrac{m_f^2}{m_\chi^2}} \, (2 m_\chi^2 + m_f^2) \bigg\{|d_6|^2 \Lambda^4 + |d_8 |^2 (2 m_\chi^2 + m_f^2)^2 - \Lambda^2 \bigg(d_6 d_8^* + d_6^* d_8 \bigg) (2 m_\chi^2 - m_f^2) \bigg\} \, ,
\label{eq:cross section two body final state}
\end{align}
where $m_\chi$ and $m_f$ denote the mass of the dark matter particle and Standard Model fermion respectively.  From Eq.~\eqref{eq:cross section two body final state}, one can see that due to the $v^2$ dependence of $\sigma v$, constraints from dark matter annihilation in the present epoch will be very weak.  Having an additional vector boson in the final state removes this $v^2$ dependence at leading order~\cite{DeSimone:2013gj,Ciafaloni:2011sa,Bell:2011if,Bell:2011eu,Bell:2010ei}.

\subsection{$\chi + \chi \rightarrow f + \bar{f} + \gamma$}

We now calculate the cross section for the process $\chi (k_1) + \chi (k_2) \rightarrow f (p_1) + \bar{f} (p_2) + \gamma (k)$.  For completeness we will show the explicit steps in our calculation.  From Eq.~\eqref{eq:dim8}, the connection between the dark matter particle and the vector bosons come from the covariant derivative: $D_\mu \equiv \partial_\mu - i g W_\mu ^i (\sigma^i/2) - i g' Y_f B_\mu$.  This can also be written as $D_\mu \equiv \partial_\mu - i (g/\sqrt{2}) (W_\mu^+ T^+ + W_\mu^- T^-) - i (g/{\rm cos}\theta_W) Z_\mu (T_3 - {\rm sin}^2 \theta_W Q) - i e Q \, A_\mu$.  To derive this, we use $W_\mu^\pm = (1/\sqrt{2})(W_\mu^1 \mp i W_\mu^2)$, $T^i = \sigma^i/2$, $T^\pm = T^1 \pm i T^2$, $Z_\mu^0 = {\rm cos} \theta_W W_\mu ^3 - {\rm sin} \theta_W B_\mu$, $A_\mu = {\rm sin} \theta_W W_\mu^3 + {\rm cos} \theta_W B_\mu$, and ${\rm cos} \theta_W = g/(\sqrt{g^2 + g'^2})$.  We denote the photon by $A_\mu$ and the weak mixing angle by $\theta_W$.  

\begin{figure*}[t]
\centering
\includegraphics[angle=0.0,width=0.25\textwidth]{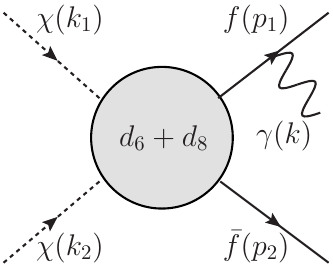}
\includegraphics[angle=0.0,width=0.25\textwidth]{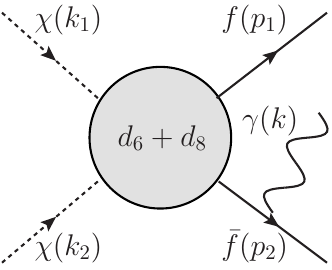}
\includegraphics[angle=0.0,width=0.25\textwidth]{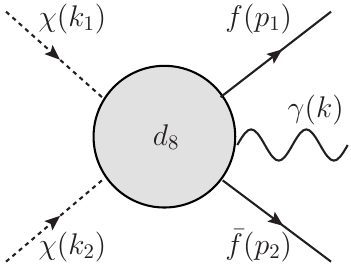}
\caption{Feynman diagrams for dark matter annihilation into a fermion - anti-fermion pair and a photon.  The dark matter, fermion, anti-fermion and the photon are denoted by $\chi$, $f$, $\bar{f}$ and $\gamma$ respectively.  The four-momentum associated with each particle is given in parenthesis next to the particle.  The operator which contributes to the Feynman diagram is written in the blob.}
\label{fig:Feynman diagrams -- momentum directions shown}
\end{figure*}

The effective Lagrangian for the given process is 
\begin{eqnarray}
\mathcal{L}_{\rm eff} &=& \dfrac{d_6}{\Lambda^2} (\bar{\chi} \gamma^5 \gamma^\mu \chi)(\bar{f} \gamma_\mu f) + \dfrac{d_8}{\Lambda^4} (\bar{\chi} \gamma^5 \gamma^\nu \chi) \bigg[(\partial_\rho \bar{f})\gamma_\nu (\partial^\rho f) \nonumber\\
&+& i \, e \, Q \, A_\rho \bigg\{(\partial^\rho \bar{f}) \gamma_\nu f - \bar{f} \gamma_\nu (\partial^\rho f)\bigg\} \bigg] \, .
\label{eq:terms relevant for f fbar gamma}
\end{eqnarray}
The relevant Feynman diagrams are shown in Fig.~\ref{fig:Feynman diagrams -- momentum directions shown}.

\begin{figure}[t]
\includegraphics[scale=0.45]{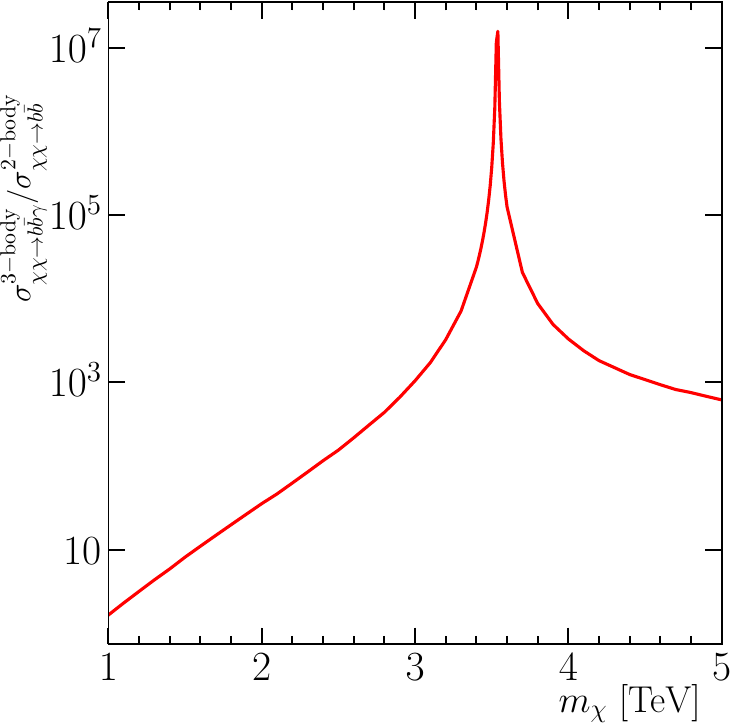}
\includegraphics[scale=0.45]{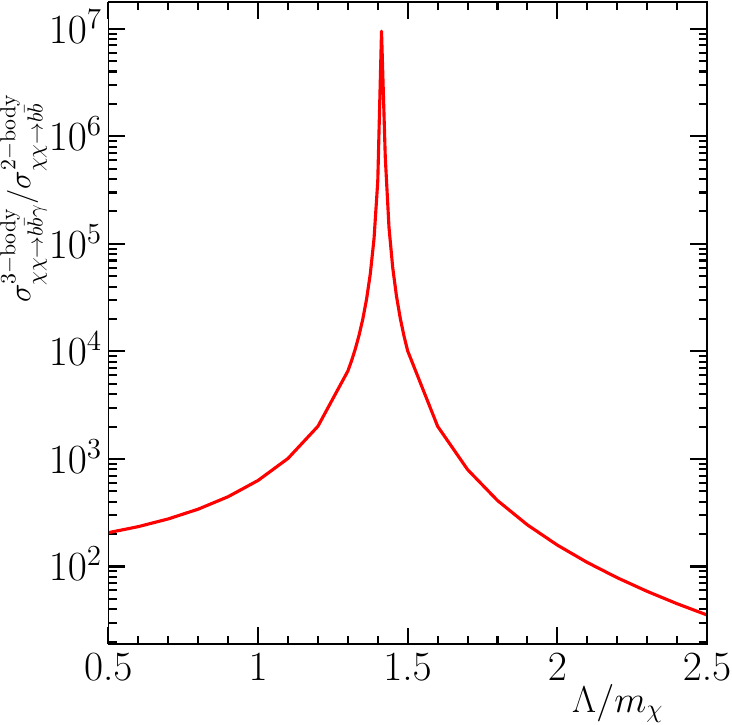}
\includegraphics[scale=0.45]{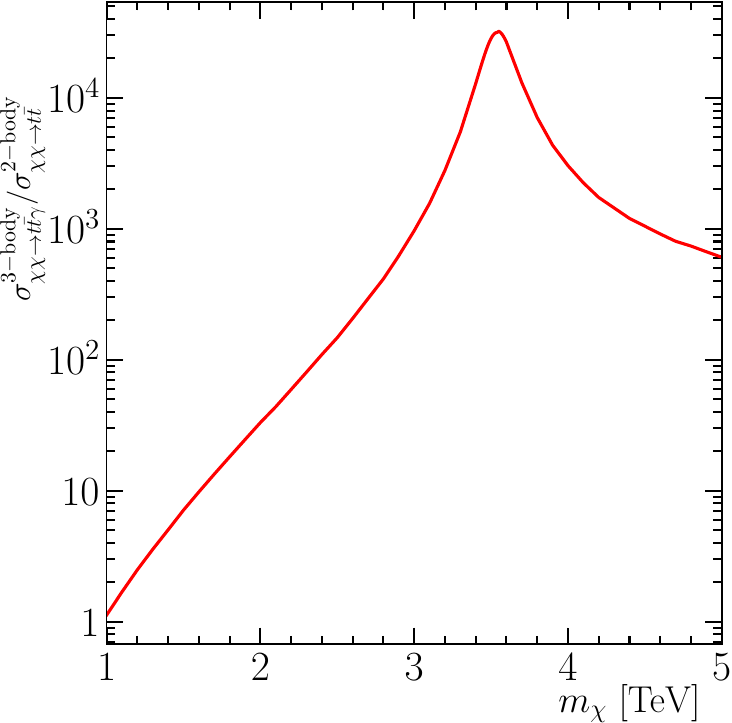}
\includegraphics[scale=0.45]{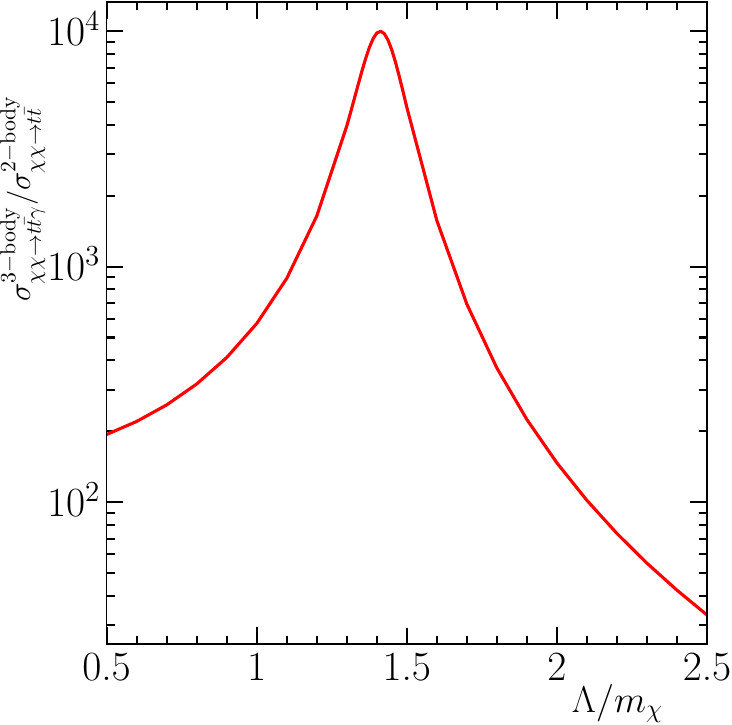}
\caption{({\bf 1$^{st}$ column:}) Ratio of 3-body cross section ($\chi \chi \rightarrow b \bar{b} \gamma$) to the 2-body cross section ($\chi \chi \rightarrow b \bar{b}$) as a function of the dark matter mass, $m_\chi$ (left), and as a function of the ratio of the EFT scale to the dark matter mass, $\Lambda/ m_\chi$ (right). In the left panel we set $\Lambda = 5$ TeV and in the right panel we set $m_\chi = 2$ TeV.  We set $d_6 = 1$, $d_8 = 1$, $v = 10^{-3}c$ for both these plots.  ({\bf 2$^{nd}$ column:}) The same for the final state comprising of top quarks.}
\label{fig:xs_ratio}
\end{figure}

\begin{figure}[b]
\includegraphics[scale=0.45]{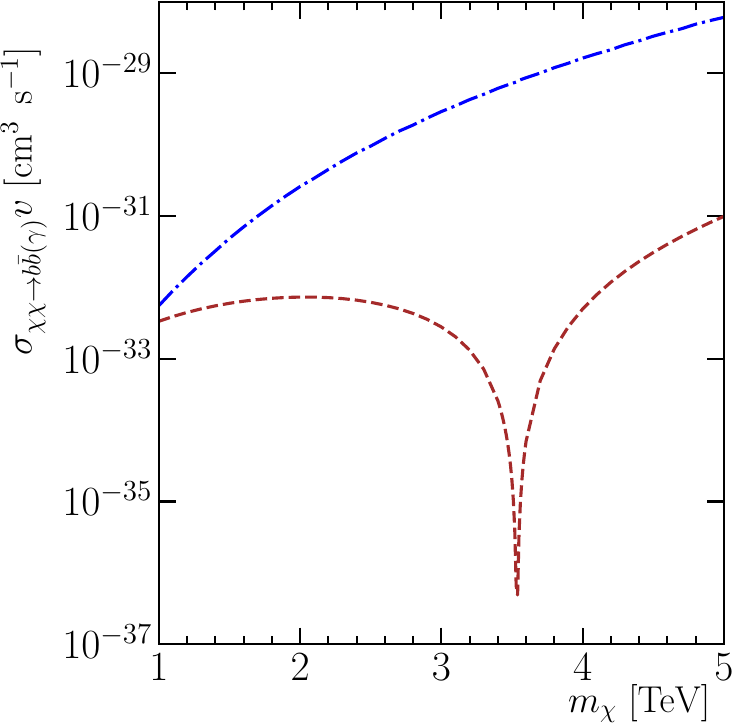}
\includegraphics[scale=0.45]{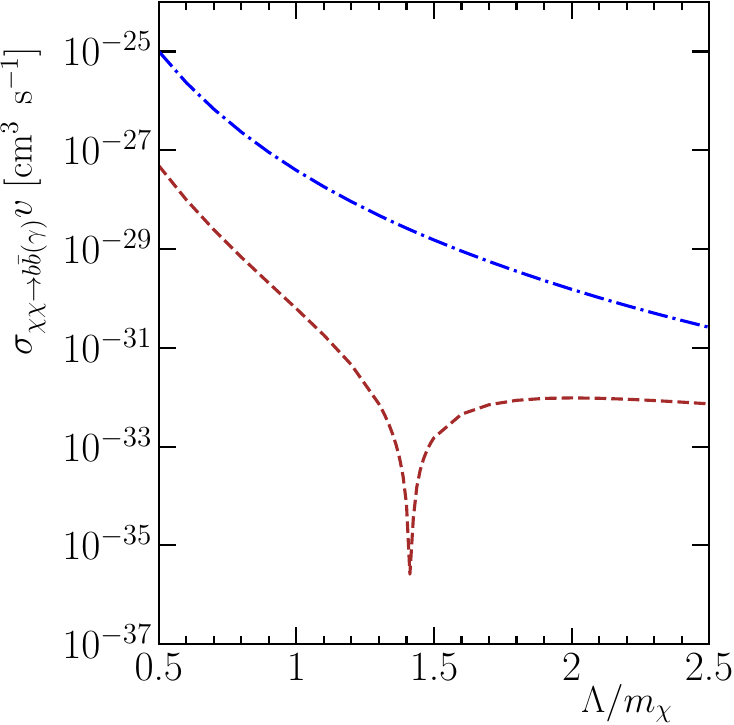}
\includegraphics[scale=0.45]{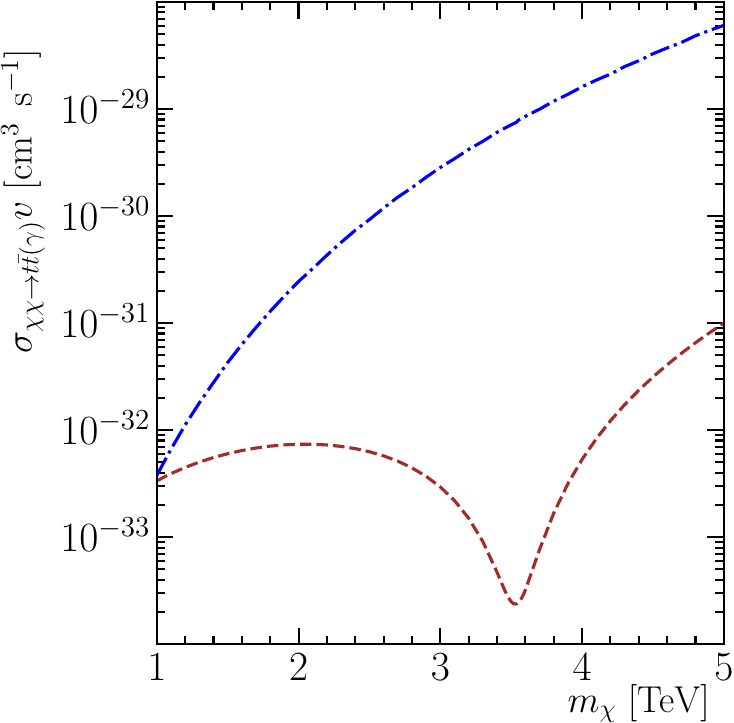}
\includegraphics[scale=0.45]{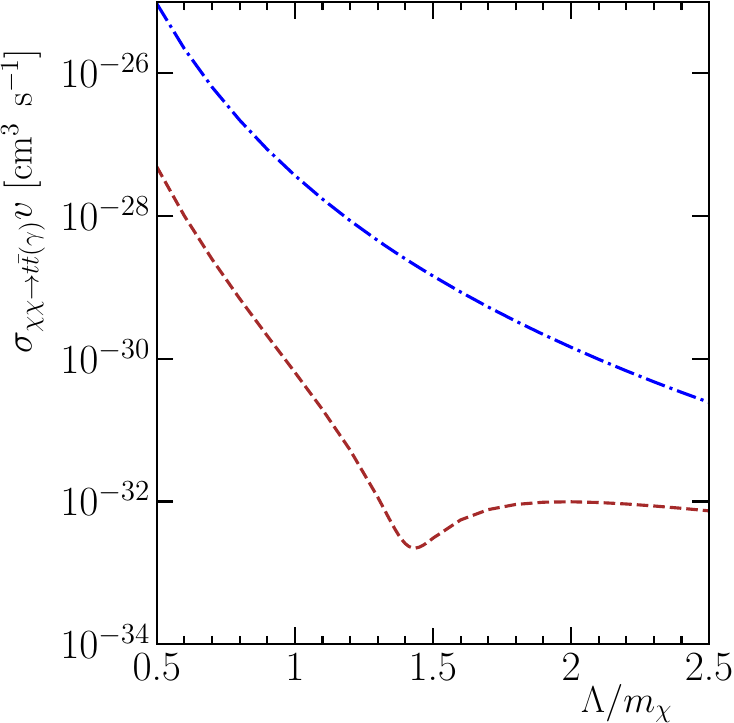}
\caption{({\bf 1$^{st}$ column:}) cross sections for the 3-body and 2-body process as a function of the dark matter mass, $m_\chi$ (left), and as a function of the ratio $\Lambda/ m_\chi$ (right).  In the left panel we set $\Lambda = 5$ TeV and in the right panel $m_{\chi} = 2$ TeV.  The 3-body cross section is denoted by the blue dash-dotted line and the brown dashed line denotes the 2-body cross section.  We set $d_6 = 1$, $d_8 = 1$, and $v = 10^{-3}c$ for both of these plots.  ({\bf 2$^{nd}$ column:}) The same for the final state comprising of top quarks.}
\label{fig:xs_ratio-new}
\end{figure}

\begin{figure}[t]
\includegraphics[scale=0.4]{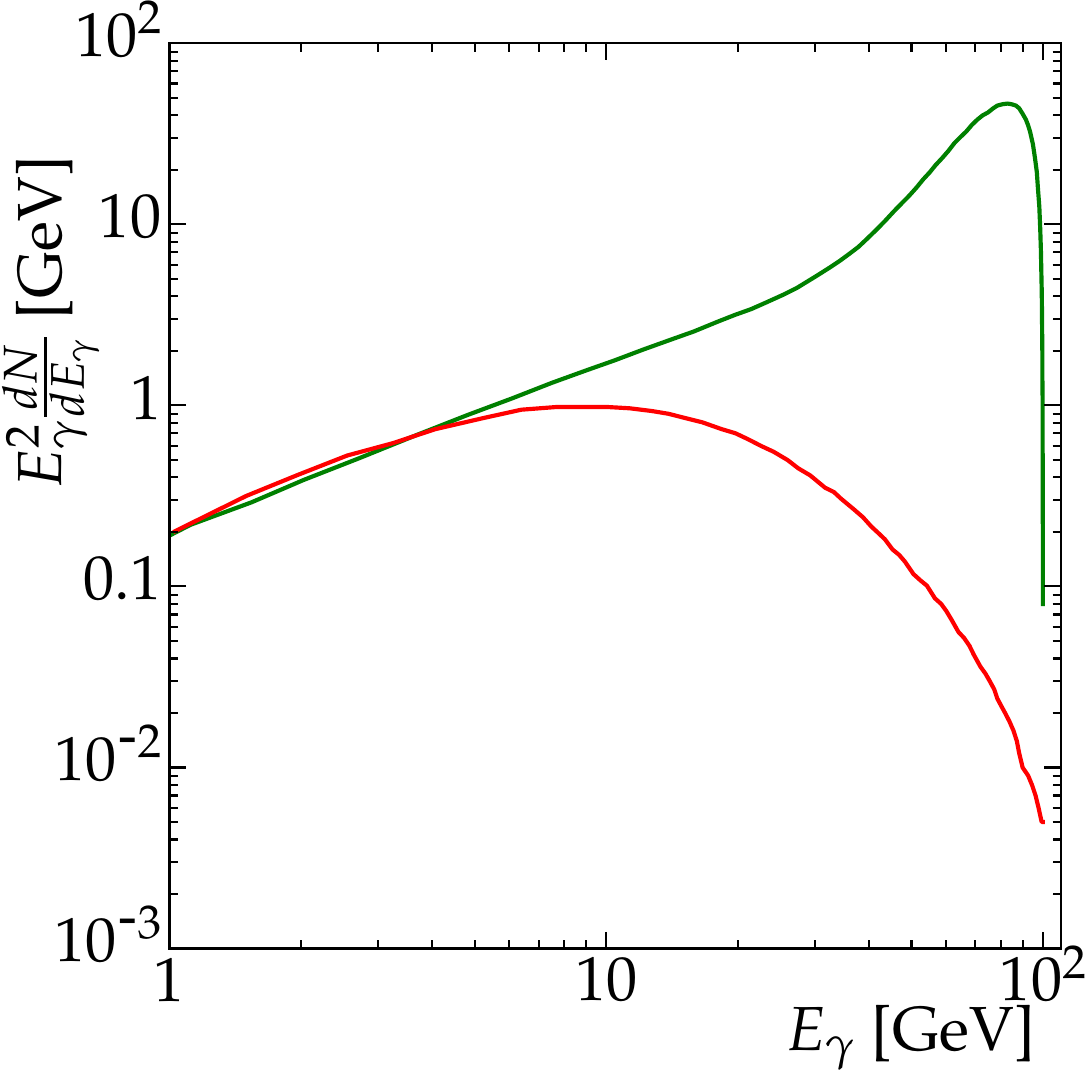}
\includegraphics[scale=0.4]{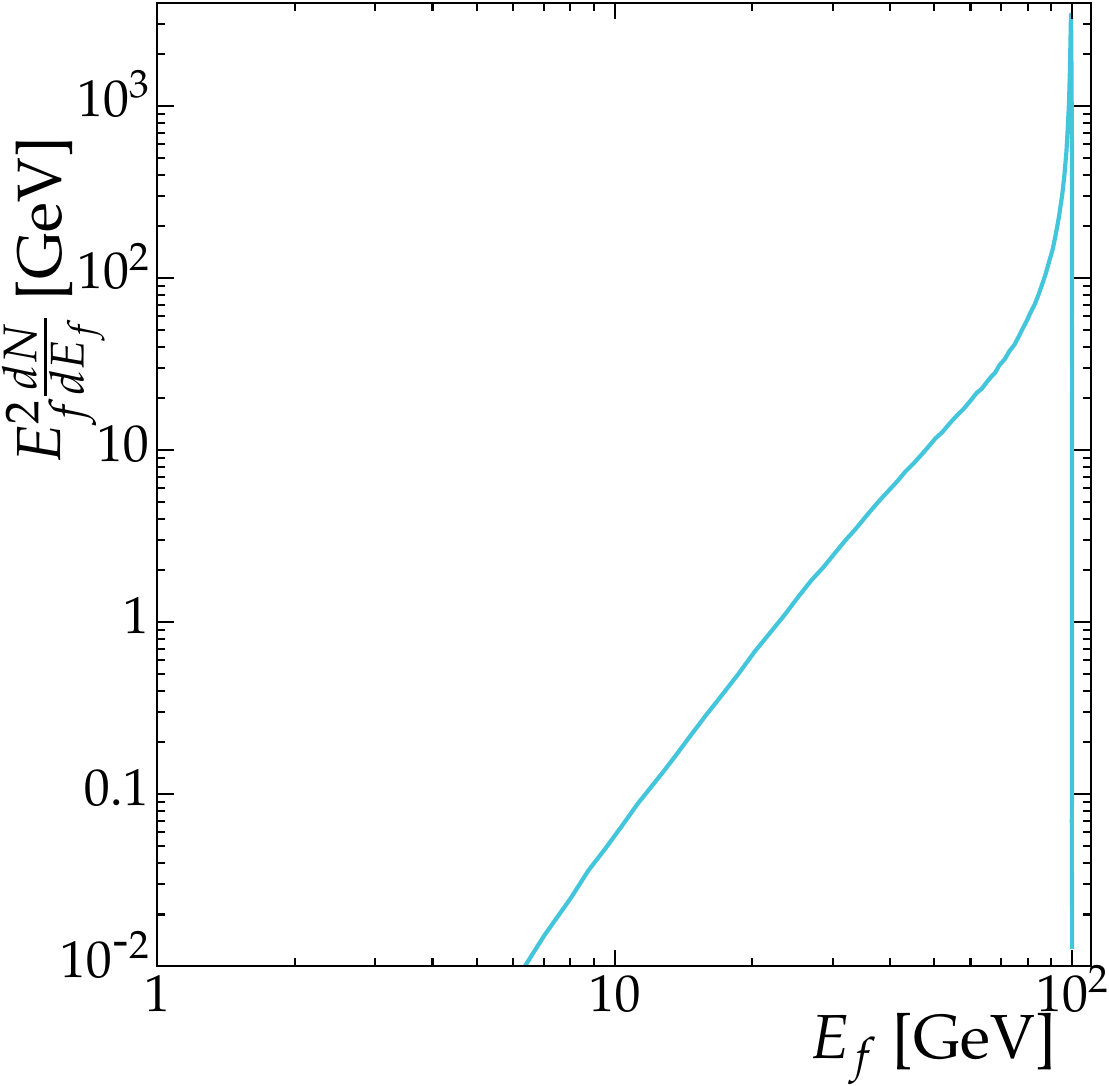}
\caption{Energy spectrum of the prompt $\gamma$-ray (left) and the bottom-quark (right) for the process $\chi \chi \to b \bar{b} \gamma$. The continuous red-line on the top panel denotes the $\gamma$-ray spectrum from the $b$-quark decay. In these plots we set $d_6 = 1$, $d_8 = 1$, $v = 10^{-3}c$, $m_\chi = 100$ GeV, and $\Lambda = 1$ TeV.}
\label{fig:energy_spectrum} 
\end{figure} 

\begin{figure}[t]
\includegraphics[scale=0.45]{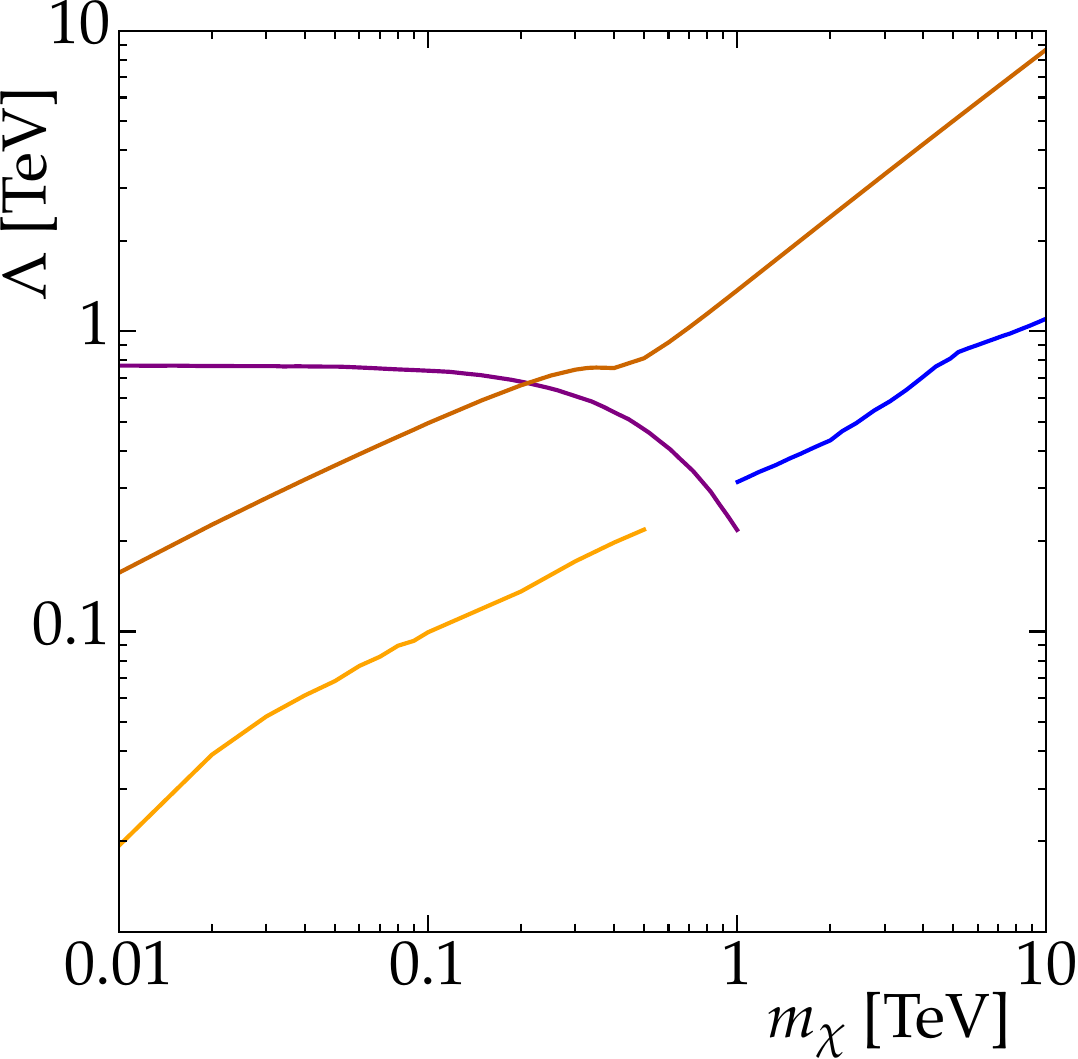}
\caption{Lower bound on the effective operator scale $\Lambda$ as a function of the dark matter mass $m_\chi$ from indirect detection experiments and collider experiments.  The orange and blue lines denote the bound derived from $\gamma$-ray fluxes from Fermi-LAT \cite{Ackermann:2015zua}, and MAGIC \cite{Aleksic:2013xea} experiment.  The dark matter relic density, as measured by the Planck collaboration~ \cite{Ade:2015xua}, is satisfied when $\Lambda$ lies on the brown line. The purple line is the bound coming from mono-$\gamma$ searches at the collider \cite{Cheung:2012gi} for the operator $\mathcal{L}_{d=6}$.  If $\Lambda$ lies below the brown line then dark matter particles make up a fraction of the relic density using the 3-body annihilation channel.  We set $d_6 = d_8 = 1$.}
\label{fig:lambda_bound}
\end{figure}

The amplitude for the process is given by
%\begin{widetext}
\begin{eqnarray}
\mathcal{M} &=& - i \, e Q \, \bar{u}(k_1) \gamma^5 \gamma^\mu v(k_2) \bigg[ \bigg(\dfrac{d_6}{\Lambda^2} - \dfrac{d_8}{\Lambda^4} \, p_2.(p_1 + k) \bigg) \bar{u}(p_1) \slashed{\epsilon}(k) \dfrac{\slashed{p_1} + \slashed{k} + m_f}{(p_1+k)^2 - m_f^2 + i \epsilon} \gamma_\mu v(p_2) \nonumber\\
&+& \bigg(\dfrac{d_6}{\Lambda^2} - \dfrac{d_8}{ \Lambda^4} \, p_1.(p_2 + k) \bigg) \bar{u}(p_1) \gamma_\mu \dfrac{- \slashed{p_2} - \slashed{k} + m_f}{(p_2+k)^2 - m_f^2 + i \epsilon} \slashed{\epsilon}(k) v(p_2)  \nonumber\\
&-& \dfrac{d_8}{\Lambda^4} \bar{u}_{p_1} \epsilon_\rho (k) \gamma_\mu v_{p_2} (p_1^\rho - p_2^\rho)
 \bigg]\, .
\label{eq:amplitude}
\end{eqnarray}
%\end{widetext}
The amplitude contains the emission of photon from the fermion due to the operator containing the coefficients $d_6$ and $d_8$.  It also includes the emission of photon from the blob due to the operator containing the coefficient $d_8$.  The square of the amplitude includes 9 terms and we will not present it here for brevity.  The calculation of the cross section involves integration of the square of the amplitude over the 3-body phase space~\cite{Jacques:thesis,Murayama:phasespace} and this is detailed in the Appendix\,\ref{appendix}.

The differential cross section w.r.t. the fermion energy in the laboratory frame as
\begin{eqnarray}
\bigg[ \dfrac{d \, (\sigma v)}{d E_{p_1}} \bigg]_{\rm lab.} &=& \dfrac{1}{ 2 s} \dfrac{1}{4} \int \sum |\mathcal{M}|^2 \, \dfrac{d(p1_k^2)}{2^6 \, (2\pi)^4} \, \dfrac{d\phi \, d {\rm cos}\theta_P}{s} \, \times 2  \, \sqrt{4s} \, , \phantom{1111}
\label{eq:differential cross section}
\end{eqnarray}
where the meanings of the symbols are given in the Appendix.  
We can obtain the photon spectrum from the fermion as follows~\cite{Jacques:thesis}.  Let us write the energy distribution of fermion per annihilation as $\dfrac{dN_f}{d \gamma _f} = \dfrac{1}{v \sigma} \dfrac{d \, v \sigma}{d \gamma_f} $, where $\gamma_f = \dfrac{E_f}{m_f}$.  Using Pythia \cite{Sjostrand:2014zea}, we obtain the spectrum of photons from fermion decay at rest $\bigg( \dfrac{dN_\gamma}{dE} \bigg)_{\rm at \, rest}$~\cite{Sjostrand:2014zea}.  The energy distribution of photons from fermion arising from dark matter annihilation is 
\begin{eqnarray}
\bigg[ \dfrac{dN}{dE} \bigg]_{\rm lab.} &=& \int _{-1} ^{+1} \dfrac{d \, {\rm cos} \, \theta '}{2} \, \int d \gamma _f \dfrac{dN_f}{d \gamma _f} \int dE' \, \bigg( \dfrac{dN_\gamma}{dE'} \bigg)_{\rm at \, rest} \nonumber\\
&\times& \delta \big(E - [\gamma _f E' + \beta _f \gamma _f p' \, {\rm cos} \, \theta']\big)\,.  
\end{eqnarray}
Integrating over ${\rm cos} \, \theta'$, we obtain
\begin{eqnarray}
\bigg[ \dfrac{dN}{dE} \bigg]_{\rm lab.} = \dfrac{1}{2} \int _{\gamma_f = 1} ^{\gamma_{\rm max}} \dfrac{d \, \gamma _f}{\sqrt{\gamma _f^2 -1}} \, \dfrac{dN_f}{d \gamma _f} \times \int _{E'_{-}} ^{E'_+} \dfrac{dE'}{E'} \, \bigg( \dfrac{dN_\gamma}{dE'} \bigg)_{\rm at \, rest}\,,\phantom{1111} 
\end{eqnarray}
where $E'_{pm} = \gamma E \pm \gamma \beta p$ and $\gamma _{\rm max} = m_\chi/ m_f$.  We multiply by an extra factor of 1/2 as $( dN_\gamma/ dE )_{\rm at \, rest}$ is due to a fermion and anti-fermion pair at rest from Pythia and $[dN/dE]_{\rm lab.}$ is due to one fermion.  The spectrum of photons from anti-fermion is identical.
  
%%%%%%%%%%%%%%%%%%%%%%%%%%%%%%%%%%%%%%%%%%%%%%%%
\section{Numerical Results}
\label{sec:numerical results}
%%%%%%%%%%%%%%%%%%%%%%%%%%%%%%%%%%%%%%%%%%%%%%%%%%%

\subsection{Behaviour of the 2-body and 3-body cross section}

In the current section we elaborate on our analytical results.  We plot the ratio of the 3-body cross section to the 2-body cross section in Fig.~\ref{fig:xs_ratio}.  In the top left panel of Fig.~\ref{fig:xs_ratio}, we show the ratio $\sigma (\chi \chi \rightarrow b \bar{b} \gamma) / \sigma(\chi \chi \rightarrow b \bar{b})$ as a function of the dark matter mass for a fixed $\Lambda$ = 5 TeV.  We set $d_6 = 1$, $d_8 = 1$, and $v = 10^{-3}c$ for this plot.  We vary the dark matter mass from $m_\chi$ = 1 TeV to 5 TeV.  The ratio is $\lesssim$ 1 for dark matter masses less than 1 TeV.  The 3-body cross section is proportional to $\Lambda^{-8}$ and the 2-body cross section is proportional to $\Lambda^{-4}$, and at small dark matter masses, the enhancement of the bremsstrahlung is not able to overcome the suppression due to the 4 additional powers of $\Lambda$.  As $m_\chi$ increases beyond 1 TeV, the ratio starts increasing beyond 1.  The  ratio of the 3-body cross section to the 2-body cross section is $\sim$ 10 for $m_\chi \sim$ 1.5 TeV.  For this particular choice of $\Lambda$, the cross section peaks at around $m_\chi \approx$ 3.5 TeV.  The maximum value of the ratio of the 3-body cross section to the 2-body cross section is $\sim$ 2$\times 10^6$.  The ratio decreases when dark matter mass is greater than 3.5 TeV.  It is interesting to note that the ratio is asymmetric around it maximal value.  Similar considerations also apply to the ratio $\sigma (\chi \chi \rightarrow t \bar{t} \gamma) / \sigma(\chi \chi \rightarrow t \bar{t})$ which we show in the bottom left panel of Fig.~\ref{fig:xs_ratio}

In the top right panel of Fig.~\ref{fig:xs_ratio}, we show the ratio of the 3-body cross section to the 2-body cross section, $\sigma (\chi \chi \rightarrow b \bar{b} \gamma) / \sigma(\chi \chi \rightarrow b \bar{b})$, as a function of the ratio $\Lambda/m_\chi$.  We take $m_\chi = 2$ TeV and vary $\Lambda/ m_\chi$ from 0.5 to 2.5 for this plot.  The ratio peaks at $\Lambda/m_\chi$ = 1.4.  It is interesting to note that the ratio of the 3-body cross section to the 2-body cross section peaks at $\Lambda/ m_\chi = 1.4$ for both the cases that we consider.  This can be analytically understood as follows:  for the 2-body cross section there is a cancellation between the dimension 8 term and the dimension 6 term.  Setting $d_6$ = $d_8$ =1 and assuming $\Lambda \gg m_f$, we find from Eq.~\eqref{eq:cross section two body final state} that the 2-body cross section is minimized when $\Lambda/m_\chi \approx 1.4$.  A non-zero value for the 2-body cross section comes from additional corrections to this ratio which depends on the fermion mass that we consider.  Similar considerations also apply to the ratio $\sigma (\chi \chi \rightarrow t \bar{t} \gamma) / \sigma(\chi \chi \rightarrow t \bar{t})$ which we show in the bottom right panel of Fig.~\ref{fig:xs_ratio}

In Fig.~\ref{fig:xs_ratio-new} the blue (brown) dash-dotted (dashed) line corresponds to the 3-body (2-body) cross section.  The parameters assumed in the top and bottom panel of this plot are the same as that of the top and bottom panel of Fig.~\ref{fig:xs_ratio} respectively.  The cancellation between the dimension 6 term and the dimension 8 term for the 2-body annihilation channel $\chi \chi \rightarrow b \bar{b}$ is clearly seen in this plot.  The top panel shows that the cross section for the 3-body annihilation channel $\chi \chi \rightarrow b \bar{b} \gamma$ monotonically increases with increasing $m_\chi$ for a fixed $\Lambda$.  The bottom panel shows that the cross section for the 3-body annihilation channel $\chi \chi \rightarrow b \bar{b} \gamma$ monotonically decreases for an increasing $\Lambda$ when $m_\chi$ is kept fixed.

\subsection{Energy spectrum of the final state photons and fermions}

The energy spectrum of the fermion emitted at the production vertex can be obtained from the cross section of the process as follows
\begin{align}
\frac{dN_f}{d E_f} = \frac{1}{v \sigma} \dfrac{d \, (v \sigma)}{d E_f}\, ,
\end{align} 
where $E_f$ denotes the final energy of the fermion.  The energy spectrum of the photons radiated from the final state fermion is obtained using Pythia~\cite{Sjostrand:2014zea}. The events are simulated at the center-of-mass energy $2 m_{\chi}$.  In Fig.~\ref{fig:energy_spectrum} we show the energy spectrum of the prompt photon (top panel) and the final state fermion (bottom panel) for the annihilation channel $\chi \chi \rightarrow b \bar{b} \gamma$. The continuous green line in the top panel of Fig. \ref{fig:energy_spectrum} depicts the spectrum of the prompt photon, while the continuous red line shows the spectrum of the photons emitted from the final state bottom quark.  In Fig. \ref{fig:energy_spectrum}, we use the dark matter particle mass as $m_\chi = 100$ GeV, $\Lambda$ = 1 TeV, $d_6 = d_8 =1$, and $v = 10^{-3}c$.

We show the energy spectrum as $E^2 dN/dE$ in a log-log plot.  The area under the curve gives the total energy carried by the particle.  From the top panel in Fig.~\ref{fig:energy_spectrum}, we see that the prompt photon carries more energy than the energy carried by the photons coming from $b$-quark hadronisation and decay.  This can be understood intuitively as the $b$-quark hadronisation and decay produces charged anti-particles and neutrinos which also carry away a substantial portion of the energy of the $b$-quark.  

The smooth spectrum of the photons resulting from $b$-quark hadronisation and decay imply that it will be difficult to distinguish this spectrum from the spectrum of conventional astrophysical sources. On the other hand, the spectrum of the prompt photons is quite unlike anything produced in astrophysics, and it will be easier to distinguish this spectrum from the background produced by conventional astrophysical sources.

The bottom panel of Fig.~\ref{fig:energy_spectrum} shows the energy distribution of the final state $b$-quark.  We see from Fig.~\ref{fig:energy_spectrum} that most of the energy is coming from the region around $m_{\chi}$. While there is a sharp feature in the bottom-quark energy distribution, in practice it will not be visible as the bottom-quark hardronises and decays after its production. The spectrum of the decay products shows a smooth feature which is hard to discriminate from the spectrum of conventional astrophysical sources.

We do not show the spectrum of the electrons, positrons, protons, anti-protons, neutrinos and anti-neutrinos arising from $b$-quark hadronisation and decay.  They can be obtained in a similar way as we obtained the spectrum of the photons from $b$-quark hadronisation and decay. 

\subsection{Constraints on $\Lambda$}

Fig.~\ref{fig:lambda_bound} compares the constraint on the effective operator scale $\Lambda$ from various different experiments as a function of the dark matter mass.  For the indirect detection experiments, we will only consider constraints from gamma-ray experiments.  Constraints from anti-protons, and positrons are competitive, however, they are subject to additional uncertainties due to our less than precise knowledge on transport and energy loss properties of the charged anti-particles.  Constraints from neutrinos are weaker compared to the upper limits from gamma-ray, and charged particle experiments.

Gamma-ray experiments typically only present the constraints on 2-body annihilation processes.  However, Fermi-LAT and MAGIC experiment presented the bin-by-bin upper limit from their experiments  in Refs.~\cite{Ackermann:2015zua} and \cite{Aleksic:2013xea} respectively.  We compare these bin-by-bin upper limits with our derived total gamma-ray spectrum to derive the constraints on $\Lambda$.

The orange and blue line is the bound obtained from $\gamma$-ray fluxes from Fermi-LAT \cite{Ackermann:2015zua} and MAGIC \cite{Aleksic:2013xea} experiment respectively. For a given $m_\chi$ the region below these lines will have lower $\Lambda$ and will produce much more $\gamma$-ray fluxes than the observed one, thus they are disfavored. While the relic density is inversely proportional to the cross section, and thus it is proportional to some power of the effective operator scale. So, the region above the brown line will have large relic abundance of the dark matter which is in contradiction to the Planck collaboration \cite{Ade:2015xua} results. For a given $m_\chi$, on the brown line one satisfies the relic density, whereas in the region below the brown line relic density is lower than the Planck measurement and to address the Planck measured relic abundance, here one needs to invoke the idea of multi-component dark matter or non-thermal production of the the dark matter.  

To compare this limit as obtained from the indirect detection experiments, with the direct detection experiments like the LHC, we show the purple line in Fig. \ref{fig:lambda_bound}. The purple line denotes the lower bound on the $\Lambda$ from the mono-$\gamma$ searches at the LHC for $d=6$ operator only (see Eq.\eqref{eq:dim6}) \cite{Cheung:2012gi}.  Thus the region below the purple line is disfavored from the mono-$\gamma$ searches at the collider. This is in contradiction to the relic density limit until the mono-$\gamma$ search limit crosses the relic density limit at $m_\chi ~ 200$ GeV or so. Hence we can see that above $m_\chi > 200$ GeV there is a common allowed region which is safe from both indirect and collider searches.

The validity of the effective field theory paradigm depends on the details of the coupling constants~\cite{Cheung:2010ua,Carpenter:2015xaa}.  In general, one can take it to be $m_\chi \lesssim \Lambda$, where the inequality can change by a factor of few depending on the underlying UV-complete model.  The value of $\Lambda$ that we derive when we assume that the 3-body annihilation process makes up the full relic density is consistent with the regime of validity.  The indirect detection limits that we present are in a regime in which one can question the validity of the effective field theory.  This implies that if one sees a signal in the very near future in indirect detection experiments, then the effective field theory approach will not be a good description of the signal for the model that we have considered.

%%%%%%%%%%%%%%%%%%%%%%%%%%%%%%%%%%%%%%%%%%%%%%%%%%%%%%%%%%%%%%
\section{Conclusions}
\label{sec:conclusions}
%%%%%%%%%%%%%%%%%%%%%%%%%%%%%%%%%%%%%%%%%%%%%%%%%%%%%%%%%%%%%%

In this paper we have considered the annihilation of two Majorana dark matter particles into SM light fermions. The annihilation of self-conjugate dark matter particles to a pair of fermions is helicity suppressed, but a photon radiation lifts up this suppression. This opens up the $s$-wave channel irrespective of the dark matter relative velocity. We addressed this phenomena from an effective operator point of view, by using a dimension 8 operator.

We have calculated the cross section of the dark matter annihilation to light fermions and a photon in the presence of dimension 6 and dimension 8 operator. We have found that in spite of the higher dimensionality of the dimension 8 operator, it does not suffer from any suppression due to dark matter relative velocity. We have showed that the contribution to the annihilation cross section from dimension 8 operator to the dimension 6 operator is always larger at all dark matter mass scales $\gtrsim 1$ TeV. Also, we have shown that there is a cancellation in the 2-body cross section between the pure dimension 6 and dimension 8 to their interference terms for $d_6 = 1$, $d_8 = 1$, and $\Lambda/m_{\chi} \simeq 1.4$. 

We have shown that the photons from VIB receives dominant contribution from prompt decay rather than the decay of the final state fermions. We then calculated the bounds on the EFT scale as a function of the dark matter mass from various indirect detection experiments like,  $\gamma$-ray fluxes measured at the Fermi-LAT and MAGIC experiment and the dark matter relic density measured by the Planck collaboration. We have compared these bounds with the mono-$\gamma$ searches at the collider.  We have found that for low dark matter mass ($\lesssim 200$ GeV) the mono-$\gamma$ searches supersedes the bound coming from the indirect detection experiments.  We expect the indirect detection bounds will improve for dark matter masses $\gtrsim$ 500 GeV with the advent of CTA \cite{Silverwood:2014yza,Lefranc:2016dgx,Lefranc:2016fgn}.   Whereas above 200 GeV dark matter mass the relic density constraint provides the strongest bound on the EFT scale.

\vspace{.4 cm}
%%%%%%%%%%%%%%%%%%%%%%%%%%%%%%%%%%%%%%%%
\section{Acknowledgment}
%%%%%%%%%%%%%%%%%%%%%%%%%%%%%%%%%%%%%%%%%%%%%%%%%%%%%%%%%%%%%%%%%%
We thank John F. Beacom, Thomas Jacques, Kenny C.Y. Ng, Gary Steigman, Vikram Rentala, Tuhin Roy, Andrea Thamm, and Alfredo Urbano for discussion.  The work of D.C. is supported by the European Research Council under the European Union's Seventh Framework Programme (FP/2007-2013) / ERC Grant Agreement n\textsuperscript{o} 279972 and by the Indo-French Center for Promotion of Advanced Research/CEFIPRA (Project no.~5404-2).  D.C. and R.L. have received partial support from the Munich Institute for Astro- and Particle Physics (MIAPP) of the DFG cluster of excellence ``Origin and Structure of the Universe''.  R.L. thanks KIPAC for support.  R.L. is supported by German Research  Foundation  (DFG)  under  Grant  Nos.  EXC-1098, KO 4820/1-1, FOR 2239, and from the European Research Council (ERC) under the European Union's Horizon 2020 research and innovation programme  (grant  agreement  No.  637506,  ``$\nu$Directions") awarded to Joachim Kopp.
 
%\vspace{.2cm}
%%%%%%%%%%%%%%%%%%%%%%%%%%%%%%%%%%%%%%%%%%%%%%%%%%
 \appendix
%%%%%%%%%%%%%%%%%%%%%%%%%%%%%%%%%%%%%%%%%%%%%%%%%%
\section{Phase space for $\chi \chi \rightarrow f \bar{f} \gamma$}
\label{appendix}

%%%%%%%%%%%%%%%%%%%%%%%%%%%%%%%%%%%%%%%%%%%%%%%%%%
 
In this section we detail all the steps of the phase space calculation required for the differential cross-section given in Eqn.\,\ref{eq:differential cross section}

The 3-body phase space can be written as
\begin{eqnarray}
d(3PS) &=& \dfrac{d^3 \bf{p_1}}{(2\pi)^3 2 E_1} \, \dfrac{d^3 \bf{p_2}}{(2\pi)^3 2 E_2} \, \dfrac{d^3 \bf{k}}{(2\pi)^3 2 E_\gamma} \times (2 \pi)^4 \delta^{(4)} \, (k_1 + k_2 - p_1 - p_2 - k) \, ,
\label{eq:3-body phase space formal expression}
\end{eqnarray}
where $E_1$, $E_2$, and $E_\gamma$ denote the energy component of the 4-momentum $p_1$, $p_2$, and $k$ respectively.  We simplify this phase space by decomposing it into a product of 2-body phase spaces.  Let us denote $p1_k = p_1 + k$, and $p1_k^2 = m1_k^2$.  We insert the identity $\int \dfrac{d^4 (p1_k)}{(2\pi)^4} \, (2\pi)^4 \, \delta (p1_k - p_1 - k) \, \Theta(p1_k^0) = 1$ and $\int \dfrac{d (m1_k^2)}{(2\pi)} \, (2\pi) \, \delta (p1_k^2 - m1_k^2) = 1$ in Eq.~\eqref{eq:3-body phase space formal expression}, where $p1_k^0$ is the 0$^{\rm th}$ component of the 4-vector $p1_k$.  The resulting expression can be simplified by noting that $\int \dfrac{d^4 (p1_k)}{(2\pi)^4} \, 2\pi \, \delta(p1_k^2 - m1_k^2) \, \Theta(p1_k^0) = \int \dfrac{d^3 {\bf p1_k}}{(2\pi)^3 \, 2 \sqrt{{\bf p1_k}^2 + m1_k^2}}$.  

The 3-body phase space can then be written as
\begin{eqnarray}
&& d(3PS) = \dfrac{d^3 \bf{p_1}}{(2\pi)^3 2 E_1} \, \dfrac{d^3 \bf{p_2}}{(2\pi)^3 2 E_2} \, \dfrac{d^3 \bf{k}}{(2\pi)^3 2 E_\gamma} \, \dfrac{d^3 {\bf p1_k}}{(2\pi)^3 \, 2 \sqrt{p1_k^2}} \nonumber\\
&\times& \dfrac{d(p1_k^2)}{2\pi} (2 \pi)^4 \delta^{(4)} \, (k_1 + k_2 - p1_k - p_2) 
\times (2\pi)^4 \, \delta^{(4)} (p1_k - p_1 - k)\, ,
\label{eq:3-body phase space into two 2-body phase space}
\end{eqnarray}
where $p1_k^2 = {\bf p1_k}^2 + m1_k^2$.

The three-dimensional integrals in Eq.~\eqref{eq:3-body phase space into two 2-body phase space} is Lorentz-invariant and can be calculated in any frame.  A compact expression for these can be obtained if any two of the three-dimensional integrals are integrated in the center of momentum frame of the two momentum vectors involved.
Following this strategy, we obtain the following expression in the center of momentum frame of $p1_k$:
\begin{eqnarray}
&&\dfrac{d^3 \bf{p_1}}{(2\pi)^3 2 E_1} \dfrac{d^3 \bf{k}}{(2\pi)^3 2 E_\gamma} (2\pi)^4 \delta^{(4)} (p1_k - p_1 - k) \nonumber\\ 
&=& \dfrac{d\phi_{p1_k} \, d({\rm cos} \, \theta_{p1_k})}{16 \pi^2}  
\dfrac{1}{\, 2([p_1^0 + k^0]_{p1_k})^2}\bigg[ ([p_1^0 + k^0]_{p1_k})^4 + p_1^4 \nonumber\\
&+& k^4 - 2 p_1^2 k^2 
- 2 p_1^2 ([p_1^0 + k^0]_{p1_k})^2 
- 2 k^2 ([p_1^0 + k^0]_{p1_k})^2 \bigg]^{1/2} \, ,
\label{eq:c.o.m. frame of p1_k}
\end{eqnarray}
where $\phi_{p1_k}$ and $\theta_{p1_k}$ denote the spherical polar coordinates in the center of momentum frame of $p1_k$.  The 0$^{\rm th}$ component of the 4 vectors $p_1$ and $k$ are denoted by $p_1^0$ and $k^0$.  In the center of momentum frame of $p1_k$, we have ${\bf p1_k}$ = 0, so that $p1_k^2 = (p1_k^0)^2$.  This will simplify the expression in Eq.~\eqref{eq:c.o.m. frame of p1_k} where we replace $([p_1^0 + k^0]_{p1_k})^2$ by $p1_k^2$.

Similarly we can also write 
\begin{eqnarray}
&&\dfrac{d^3 \bf{p_2}}{(2\pi)^3 2 E_2} \dfrac{d^3 (\bf{p1_k})}{(2\pi)^3 2 \sqrt{{\bf p1_k}^2+m1_k^2}} (2\pi)^4 \delta^{(4)} (P - p1_k - p_2) \nonumber\\
&&= \dfrac{d\phi_{P} \, d({\rm cos} \, \theta_{P})}{32 \pi^2 \, P^2}\,  \bigg[ P^4 + p_2^4 + (p1_k)^4 - 2 p_2^2 (p1_k)^2  - 2 p_2^2 P^2 - 2 P^2 (p1_k)^2 \bigg]^{1/2} \, ,
\label{eq:c.o.m. frame of p2 and p1k}
\end{eqnarray}
where $P = p_2 + p1_k = k_1 + k_2$.  The spherical polar coordinates in the center of momentum frame of $P$ are denoted by $\phi_P$ and $\theta_P$.  

The minimum  value of $p1_k^2 = (p_1 + k)^2$ is $m_f^2$.  We choose our coordinate system such that the integral is symmetric about $\phi_{p1_k}$.  After integrating over $\phi_{p1_k}$ we obtain
\begin{eqnarray}
d(3PS_\phi) &=& \int_{p1_k^2 = m_f^2} ^s  \left\{s^2 + p1_k^4 + m_f^4 - 2 s \, p1_k^2 - 2 s \, m_f^2 \right. \nonumber\\ 
&-& \left. 2 m_f^2 \, p1_k^2\right\}^{1/2}  \left\{p1_k^4 + m_f^4 + k^4 - 2 m_f^2 \, p1_k^2 \right. \nonumber\\
&-& \left. 2 m_f^2 \, k^2 - 2 p1_k^2 \, k^2 \right\}^{1/2} \times \dfrac{1}{2^6} \, \dfrac{1}{s \, p1_k^2} \, d\phi \, d({\rm cos} \theta_P) \, d({\rm cos} \theta_{p1_k}) \, \dfrac{d(p1_k^2)}{(2\pi)^4} \,, \phantom{1111}
\label{eq:3 dim phase space after integrating over one phi angle}
\end{eqnarray}
where we now re-denote $\phi_P$ as $\phi$.

We can now choose the individual 3-vectors as
\begin{eqnarray}
\bigg[{\bf k_1} \bigg]_P &=& \dfrac{\sqrt{s}}{2} \, |{\bf v}_\chi| \, ({\rm sin} \theta_P \, {\rm cos} \phi, \, {\rm sin} \theta_P \, {\rm sin} \phi, \, {\rm cos}\theta_P) \, ,\nonumber\\
\bigg[{\bf k_2} \bigg]_P &=& - \dfrac{\sqrt{s}}{2} \, |{\bf v}_\chi| \, ({\rm sin} \theta_P \, {\rm cos} \phi, \, {\rm sin} \theta_P \, {\rm sin} \phi, \, {\rm cos}\theta_P) \, ,\nonumber\\
\bigg[{\bf p_2} \bigg]_P &=& (0, 0, - \sqrt{(E_2^P)^2 - m_f^2}) \, ,\nonumber\\
\bigg[{\bf p1_k} \bigg]_P &=& (0, 0,  \sqrt{(E_2^P)^2 - m_f^2}) \, ,\nonumber %\\
\end{eqnarray}
\begin{eqnarray}
\bigg[{\bf p_1} \bigg]_{p1_k} &=& \sqrt{(E_1^{p1_k})^2 - m_f^2} \, ({\rm sin} \theta_q, 0, {\rm cos} \theta_q) \, ,\nonumber\\
\bigg[{\bf k} \bigg]_{p1_k} &=& - \sqrt{(E_1^{p1_k})^2 - m_f^2} \, ({\rm sin} \theta_q, 0, {\rm cos} \theta_q) \, ,
\label{explicit forms of 3-vector}
\end{eqnarray}
where $[ ... ]_P$ and $[ ... ]_{p1_k}$ denote evaluation in the laboratory and $p1_k$ rest frame respectively.  The Mandelstam variable $s = (k_1 + k_2)^2 = (p_1 + p_2 + k)^2 = (p1_k + p_2)^2$, $t_1 = (k_1 - p1_k)^2 = (k_2 - p_2)^2$.  

By definition, $p_2^0 = E_2^P$, so that $k_2 . p_2 = m_\chi \, E_2^P - m_\chi |v_\chi| \, {\rm cos} \theta_P \, \sqrt{(E_2^P)^2 - m_f^2}$.  Since $2 \, p_2 . p1_k = s - m_f^2 - p1_k^2 = 2 E_2^P p1_k^0 - 2 \bigg(- \sqrt{(E_2^P)^2 - m_f^2} \bigg) \bigg(\sqrt{(E_2^P)^2 - m_f^2} \bigg)$, and $p1_k^0 = P^0 - p_2^0 =2 m_\chi - E_2^P$, applying the expression from the second equation into the former equation we derive $E_2^P = (s + m_f^2 - p1_k^2)/(4 m_\chi)$.  This implies $t_1 = (k_1 - p1_k)^2 = (k_2 - p_2)^2 = m_\chi^2 + m_f^2 - 2 \bigg[ \dfrac{1}{4}(s + m_f^2 - p1_k^2) - m_\chi | v_\chi | \, {\rm cos} \theta_P \bigg\{\dfrac{(s + m_f^2 - p1_k^2)^2}{16 m_\chi^2} - m_f^2\bigg\}^{1/2} \bigg]$, and $u_1 = (k_1-p_2)^2 = (k_2 - p1_k)^2 =  m_\chi^2 + m_f^2 - 2 \bigg[ \dfrac{1}{4}(s + m_f^2 - p1_k^2) + m_\chi | v_\chi | \, {\rm cos} \theta_P \bigg\{\dfrac{(s + m_f^2 - p1_k^2)^2}{16 m_\chi^2} - m_f^2\bigg\}^{1/2} \bigg]$.    We have used $p_2.p1_k = 1/2 (s - m_f^2 - p1_k^2)$, $k_1 . k_2 = s/2 - m_\chi^2, \, 2 k . p_1 = p1_k^2 - k^2 - m_f^2$.  Since $p_2^2 = (p_2^0)^2 -  (E_2^P)^2 + m_f^2 = m_f^2$, we have $[p_2^0]_P = E_2^P = (s + m_f^2 - p1_k^2)/4 m_\chi$.  Similarly, from $P^0 = p1_k^0 + p_2^0$, we derive $[p1_k^0]_P = \sqrt{s}/2 \, - \, m_f^2/2 \sqrt{s} \, + \, p1_k^2/2 \sqrt{s}$.

By definition, we have $[p_1^0]_{p1_k} = E_1^{p1_k}$, $k^2 = (k^0)^2 - (E_1^{p1_k})^2 + m_f^2$, and $2 p_1 . k = p1_k^2 - m_f^2 - k^2 = 2 E_1^{p1_k} k^0 + 2 ((E_1^{p1_k})^2 - m_f^2)$.  We can solve for $k^0$ and $E_1^{p1_k}$ from the latter two equations to derive $[ k^0 ]_{p1_k} = \dfrac{p1_k^2 - m_f^2 + k^2}{2 \sqrt{p1_k^2}}$ and $[ E_1^{p1_k} ]_{p1_k} = \dfrac{p1_k^2 + m_f^2 - k^2}{2 \sqrt{p1_k^2}}$.  By definition, we have $[p1_k^\mu]_{p1_k} = (\sqrt{p1_k^2}, 0, 0, 0)$, so that the Lorentz factor for boost from the $p1_k$ rest frame to the rest frame of $P$ is $\gamma = \dfrac{[p1_k^0]_P}{[p1_k^0]_{p1_k}} =\dfrac{\sqrt{s}/2 \, - \, m_f^2/2 \sqrt{s} \, + \, p1_k^2/2 \sqrt{s}}{\sqrt{p1_k^2}}$.  This implies $[p1_k^3]_P = \sqrt{\dfrac{s}{4} + \dfrac{m_f^4}{4 s} + \dfrac{p1_k^4}{4 s} - \dfrac{m_f^2}{2} - \dfrac{p1_k^2}{2} - \dfrac{m_f^2 \, p1_k^2}{2 s}}$.

In the $p1_k$ rest frame, conservation of momentum implies $| [{\bf p_1}]_{p1_k} | = | [{\bf k}]_{p1_k}|$, from which we obtain $[E_{p_1}]_{p1_k} = \sqrt{m_f^2 + ([E_\gamma]_{p1_k})^2}$, where $[E_{p_1}]_{p1_k}$ and $[E_\gamma]_{p1_k}$ denote the energy of the particle with 4-momentum $p_1$ and the photon in the rest frame of $p1_k$ respectively.  Similarly conservation of energy in the same reference frame implies $p1_k^0 = p_1^0 + k^0 = \sqrt{p1_k^2}$, from which we obtain $E_\gamma ^{p1_k} = \dfrac{p1_k^2 - m_f^2}{2 \sqrt{p1_k^2}}$.  From all these expression, we obtain the energy of the photon in the laboratory frame as
\begin{eqnarray}
&&[E_\gamma]_P = \gamma [E_\gamma]_{p1_k} + \gamma \beta [p_\gamma ^z]_{p1_k} \nonumber\\
&=& \dfrac{(p1_k^2 - m_f^2) \bigg(\dfrac{\sqrt{s}}{2} - \dfrac{m_f^2}{2 \sqrt{s}} + \dfrac{p1_k^2}{2 \sqrt{s}} \bigg)}{2 \, q_1^2} \nonumber\\
& - & \dfrac{(p1_k^2 - m_f^2) \bigg( \dfrac{s}{4} + \dfrac{m_f^4}{4 s} + \dfrac{p1_k^4}{4 s} - \dfrac{m_f^2}{2} - \dfrac{p1_k^2}{2} - \dfrac{m_f^2 p1_k^2}{2} \bigg)}{2 \, p1_k^2} \, \times {\rm cos} \,\theta_q \, .\phantom{1111}
\label{eq:photon energy in lab frame}
\end{eqnarray}
Taking ${\rm cos} \theta_q = \pm 1$, we obtain the extremum value of $p1_k^2$.  We derive the differential cross section w.r.t. the photon energy in the laboratory frame as
\begin{eqnarray}
\bigg[ \dfrac{d \, (\sigma v)}{d E_\gamma} \bigg]_{\rm lab.} &=& \dfrac{1}{ 2 s} \dfrac{1}{4} \int \sum |\mathcal{M}|^2 \, \dfrac{d(p1_k^2)}{2^6 \, (2\pi)^4} \, \dfrac{d\phi \, d {\rm cos}\theta_P}{s} \, \times (- 2) \, \sqrt{4s} \, . \phantom{1111}
\label{eq:differential cross section appendix}
\end{eqnarray}

Similarly, in the center of momentum frame of $p1_k$, we obtain $[E_{p_1}]_{p1_k} = (p1_k^2 + m_f^2)/2\sqrt{p1_k^2}$.  The energy of fermion $f$ is
\begin{eqnarray}
&&[E_{p_1}]_P = \gamma [E_\gamma]_{p1_k} + \gamma \beta [p_\gamma ^z]_{p1_k} \nonumber\\
&=& \dfrac{(p1_k^2 + m_f^2) \bigg(\dfrac{\sqrt{s}}{2} - \dfrac{m_f^2}{2 \sqrt{s}} + \dfrac{p1_k^2}{2 \sqrt{s}} \bigg)}{2 \, q_1^2} \nonumber\\
& + & \dfrac{(p1_k^2 - m_f^2) \bigg( \dfrac{s}{4} + \dfrac{m_f^4}{4 s} + \dfrac{p1_k^4}{4 s} - \dfrac{m_f^2}{2} - \dfrac{p1_k^2}{2} - \dfrac{m_f^2 p1_k^2}{2} \bigg)}{2 \, p1_k^2} \times {\rm cos} \, \theta_q \,.
\label{eq:fermion energy in lab frame}
\end{eqnarray} 

\bibliographystyle{JHEP}
\bibliography{Effective_operator_JHEP_v1}
\end{document}